\documentclass[fleqn,12pt]{wlscirep}
\usepackage{lineno}
\usepackage{setspace}
\usepackage{algorithm2e}
\usepackage{float}
\usepackage[hyphens]{url}

\title{On the predictability of infectious disease outbreaks}

\author[1,2,3,4,5$\dagger$,*]{Samuel V. Scarpino}
\author[5,6,$\dagger$,**]{Giovanni Petri}
\affil[1]{Network Science Institute, Northeastern University, Boston, MA, 02115, USA}
\affil[2]{Marine \& Environmental Sciences, Northeastern University, Boston, MA, 02115, USA}
\affil[3]{Physics, Northeastern University, Boston, MA, 02115, USA}
\affil[4]{Health Sciences, Northeastern University, Boston, MA, 02115, USA}
\affil[5]{ISI Foundation, 10126 Turin, Italy}
\affil[6]{ISI Global Science Foundation, New York, NY 10018, USA}
\affil[$\dagger$]{Both authors contributed equally to this work.}
\affil[*]{s.scarpino@northeastern.edu}
\affil[**]{giovanni.petri@isi.it}

\begin{document}
\maketitle
\thispagestyle{empty}
\doublespacing 
\flushbottom
\textbf{Infectious disease outbreaks recapitulate biology: they emerge from the multi-level interaction of hosts, pathogens, and their shared environment.  As a result, predicting when, where, and how far diseases will spread requires a complex systems approach to modeling.  Recent studies have demonstrated that predicting different components of outbreaks--e.g., the expected number of cases, pace and tempo of cases needing treatment, demand for prophylactic equipment, importation probability etc.--is feasible~\cite{colizza2007predictability, shaman:RealTime, venkatramanan2017using, johansson2016evaluating, brooks2015flexible, funk2016real, chowell2016using, zhang2015social, nsoesie2013simulation}.  Therefore, advancing both the science and practice of disease forecasting now requires testing for the presence of fundamental limits to outbreak prediction~\cite{chretien2015advancing,biggerstaff2016results,ostpForecasting, gandon2016forecasting}.  To investigate the question of outbreak prediction, we study the information theoretic limits to forecasting across a broad set of infectious diseases using permutation entropy as a model independent measure of predictability~\cite{garland2014model}.  Studying the predictability of a diverse collection of historical outbreaks--including, chlamydia, dengue, gonorrhea, hepatitis A, influenza, measles, mumps, polio, and whooping cough--we identify a fundamental entropy barrier for infectious disease time series forecasting.  However, we find that for most diseases this barrier to prediction is often well beyond the time scale of single outbreaks, implying prediction is likely to succeed.  We also find that the forecast horizon varies by disease and demonstrate that both shifting model structures and social network heterogeneity are the most likely mechanisms for the observed differences in predictability across contagions.  Our results highlight the importance of moving beyond time series forecasting, by embracing dynamic modeling approaches to prediction~\cite{colizza2006role}, and suggest challenges for performing model selection across long disease time series.  We further anticipate that our findings will contribute to the rapidly growing field of epidemiological forecasting and may relate more broadly to the predictability of complex adaptive systems.}


``If we don’t have a vaccine—yes, we are all going to get it.~\cite{shaw2007sars}" This dire assessment by a Canadian nurse in 2003 reflected the global public health community's worst fears about the ongoing SARS outbreak~\cite{dye2003modeling, colizza2007predictability}. These fears--for perhaps the first time in history--were partially derived from mathematical and computational models, which were developed in near real-time during the outbreak to forecast transmission risk~\cite{chretien2015advancing, colizza2007predictability}.  However, the predictions for SARS failed to match the data~\cite{meyers2005network, colizza2007predictability}.  Over the subsequent fifteen years, the scientific community developed a rich understanding for how social contact networks, variation in health care infrastructure, the spatial distribution of prior immunity, etc., drive complex patterns of disease transmission~\cite{perra2015modeling, gandon2016forecasting, reich2016challenges, viboud2018rapidd, peak2018population, wesolowski2015impact} and demonstrated that data-driven, dynamic and\slash or agent-based models can produce actionable forecasts~\cite{bansal2016big, funk2016real, pastore2016real, lofgren2014opinion, chowell2017perspectives, ray2018prediction}. Despite these advances, an ongoing debate continues in the scientific community about both the need and our capacity to forecast outbreaks~\cite{holmes2018pandemics, rivers2018modelling}. What remains an open question is whether the existing barriers to forecasting stem from gaps in our mechanistic understanding of disease transmission and low-quality data or from fundamental limits to the predictability of complex, sociobiological systems, i.e. outbreaks~\cite{hufnagel2004forecast, perra2015modeling, moran2016epidemic, gandon2016forecasting}.

\begin{figure}[t]
\centering
\includegraphics[width=0.99\linewidth]{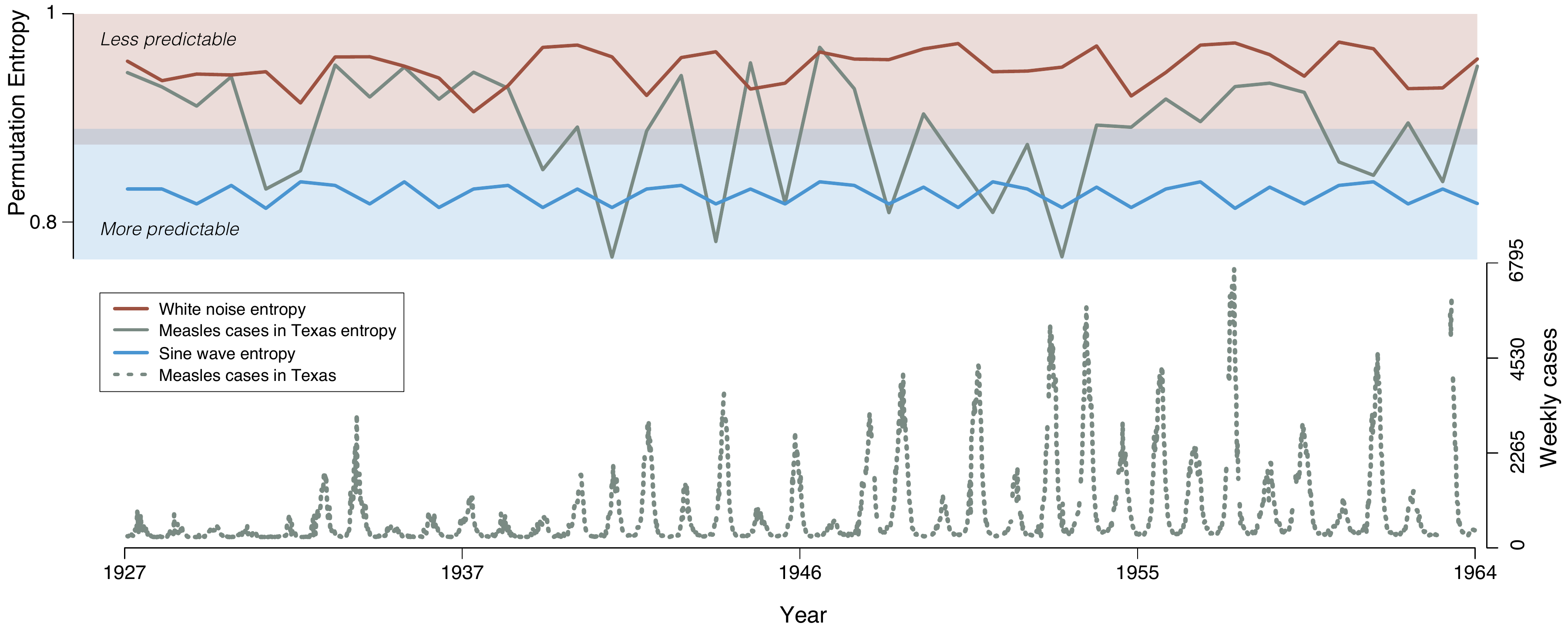}
\caption{\label{fig:timeserie}\textbf{Permutation entropy varies through time for real-world disease time series.} The permutation entropy in windows of size 52 weeks across three time series of equal length: \textit{1.) dark red:} Gaussian--white--noise ($\mu = 0, \sigma = 1$); \textit{2.) gray:} Measles cases reported to the CDC from the state of Texas between 1927 and 1965 (gray dashed line, lower panel) as digitized from MMWR reports by Project Tycho~\cite{tycho}; and \textit{3.) blue:} a sine wave with Gaussian noise ($\mu = 0, \sigma = 0.01$).  The fluctuations in the permutation entropy calculated from the measles time series (gray dashed line) are larger than would be predicted by chance and result in periods of time where model-based forecasts should be accurate (in the range of a noisy sine wave, blue shaded region) and periods of time (in the range of the white noise, red shaded region) when statistical forecasts based solely on the time series data should outperform model-based forecasts.}
\end{figure}

In order to study the predictability of diseases in a comparative framework, which also permits stochasticity and model non-stationarity, we employ permutation entropy as a model-free measure of time series predictability ~\cite{bandt2002permutation,garland2014model,pennekamp2018intrinsic}.
This measure, i.e permutation entropy, is ideal because--in addition to being a model independent metric of predictability--recent work has demonstrated that it correlates strongly with known limits to forecasting in dynamical systems, e.g., models where we can measure Lyapunov stability~\cite{bandt2002permutation,garland2014model,pennekamp2018intrinsic} and can be transformed into an estimate of Kolmogorov-Sinai entropy~\cite{politi2017quantifying}.  Additionally, a recent study by Pennekamp et al. (2018) demonstrated that permutation entropy correlated strongly with forecast accuracy for ecological models~\cite{pennekamp2018intrinsic}.

Permutation entropy is conceptually similar to the well-known Shannon entropy~\cite{bandt2002permutation}.  However, instead of being based on the probability of observing a system in a particular state, it utilizes the frequency of discrete motifs, i.e symbols, associated with the growth, decay, and stasis of a time series.  For example, in a binary time series the permutation entropy in two dimensions would count the frequency of the set of possible ordered pairs, \{[01],[10]\}, and the Shannon entropy, or uniformity, of this distribution is the permutation entropy.  In higher dimensions, one can define an alphabet of symbols over all factorial combinations of orderings in a given dimension, e.g., \{[0, 1, 2], [2, 1, 0], [1, 0, 2], etc.\}, over which the permutation entropy will be defined.  A time series that visits all the possible symbols with equal frequency will have maximal entropy and minimal predictability and a time series that only samples a few of the possible symbols will instead have lower entropy and hence be more predictable.

More formally, for a given time series $\{x_ t \}_{t =1, \ldots, N}$ indexed by positive integers, an embedding dimension $d$ and a temporal delay $\tau$, we consider the set of all sequences of value $s$ of the type $s = \{ x_t, x_{t+\tau}, \ldots, x_{t+ (d-1)\tau} \}$. 
To each $s$, we then associate the permutation $\pi$ of order $d$ that makes $s$ totally ordered, that is $\tilde{s} = \pi(s) = [x_{t_i}, \ldots,x_{t_N}]$ such that $x_{t_i} < x_{t_j} \, \forall t_i \, < t_j$, hence generating the symbolic alphabet. Ties in neighboring values, i.e. $x_{t_i} = x_{t_j}$, were broken both by keeping them in their original order in the time series and\slash or by adding a small amount of noise, the method of tie breaking did not affect the results, see~\cite{zunino2017permutation} for more details on tie-breaking and permutation entropy. The permutation entropy of time series $\{x_t\}$ is then given by the Shannon entropy on the permutation orders, that is $H^p_{d,\tau} (\{ x_t\}) = - \sum_\pi p_\pi \log p_\pi$, where $p_\pi$ is the probability of encountering the pattern associated with permutation $\pi$. 

As described above, calculating the permutation entropy of a time series requires selecting values for the embedding dimension $d$, the time delay $\tau$, and the window length $N$. In this study, our goal was to find conservative values of $H^p$ by searching over a wide range of possible $(d,\tau)$ pairs and setting $H^p (\{ x_t\}) = \min_{d,\tau} H^p_{d,\tau}(\{ x_t\})$. However, the value of $H^p$ should always decline as the embedding dimension $d$ grows, i.e. no minimum value of $H^p$ will exist for finite windows sizes $N$.  To address this issue, we follow Brandmaier (2015) and exclude all unobserved symbols when calculating $H^p$, which acts as a ``penalty'' against higher dimensions and results in a minimum value of $H^p$ for finite length time series.  To control for differences in dimension and for the effect of time series length on the entropy estimation, we normalize the entropy by log$(d!)$, ensure that each window is greater in length than $d!$, and confirm that the estimate of $H^p$ has stabilized (specifically that the marginal change in $H^p$ as data are added is less than $1\%$). To facilitate interpretation, we present results from continuous intervals by fixing $\tau=1$.  However, our results generalize to the case where we fix both $d$ and $\tau$ across all diseases and where we minimize over a range of $(d,\tau)$ pairs (see Supplement).

Permutation entropy does not require the \textit{a priori} specification of a mechanistic nor generating model, which allows us to study the predictability of -potentially very different- systems within a unified framework. What is not explicit in the above formulation is that the permutation entropy can be accurately measured with far shorter time series than Lyapunov exponents and that it is robust to both stochasticity and linear\slash nonlinear monotonous transformations of the data, i.e. it is equivalent for time series with different magnitudes~\cite{bandt2002permutation,zunino2012distinguishing}. Consider--for example--two opposite cases with respect to their known predictability, pure white noise and a perfectly periodic signal. We expect the former, being essentially random, to display a very high entropy as compared with the latter, which instead we expect to show a rather low entropy in consideration of its simple periodic structure. 

In Figure \ref{fig:timeserie}, we demonstrate that this is indeed the case, even when we allow the periodic signal to be corrupted by a small amount of noise. We track the short scale predictability of the time series by calculating the permutation entropy in moving windows (with width = one year, although the results are robust to variation in window size). For comparison, we calculate the same moving window estimate of the permutation entropy for the time series of measles cases in Texas prior to the introduction of the first vaccine.  
The critical observation is that the moving-window entropy for the measles time series fluctuates between values comparable with that of pure random noise and, at times, values closer to the more predictable periodic signal, which suggests alternating intervals with different dynamical regimes and, thus, predictability. The magnitude of the entropy fluctuations for measles in Texas is statistically significant by permutation test, $p<0.001$, as compared with simulated fluctuations obtained by building an estimated multinomial distribution over the symbols and repeatedly calculating the expected Jenson-Shannon divergence from simulations.

\begin{figure}[H]
\centering
\includegraphics[width=0.73\linewidth]{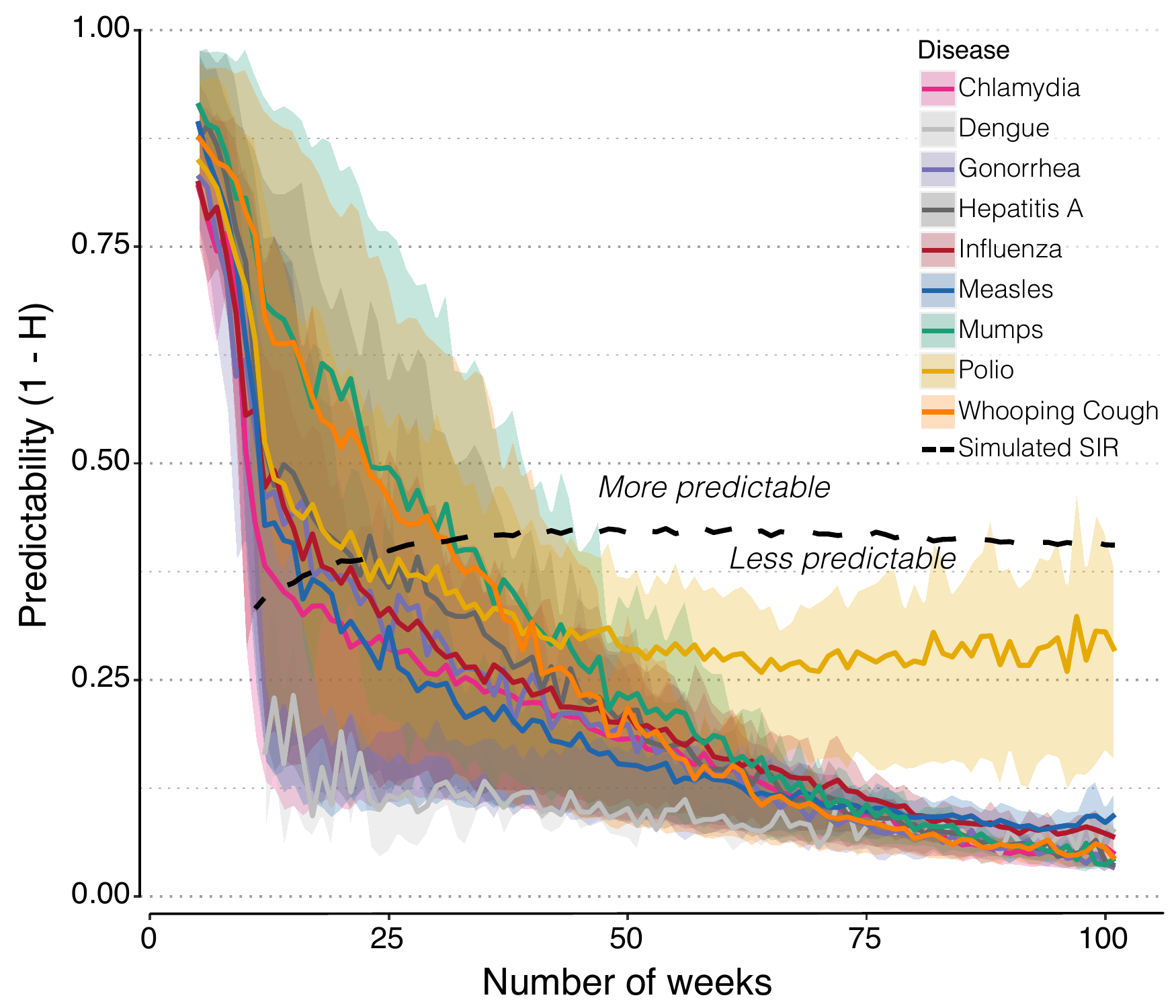}
\caption{\label{fig:smallrambo}\textbf{Single outbreaks are often predictable.} The average predictability ($1 - H^p$) for weekly, state-level data from nine diseases is plotted as a function of time series length in weeks.  For each disease, we selected 1,000 random starting locations in each time series and calculated the permutation entropy in rolling windows in lengths ranging from 2 to 104 weeks.  The solid lines indicate the mean value and the shaded region marks the interquartile range across all states and starting locations in the time series.  Although the slopes are different for each disease, in all cases, longer time series result in lower predictability.  However, most diseases are predictable across single outbreaks and disease time series cluster together, i.e. there are disease-specific slopes on the relationship between predictability and time series length.  To aid in interpretation, the black line plots the median permutation entropy across 20,000 stochastic simulations of a Susceptible Infectious Recovered (SIR) model, as described in the Supplement.  This SIR model would be considered ``predictable,'' thus values above the black-line might be thought of as in-the-range where model-based forecasts are expected to outperform forecasts based solely on statistical properties of the time series data.}
\end{figure}

We now turn our attention to a broader set of diseases and ask how the predictability, defined as $\chi = 1 - H^p$ (where $H^p$ is the permutation entropy), scales with the amount of available data (i.e. the time series' length).  Specifically, we compute the permutation entropy across more than 25 years of weekly data at the US state-level for chlamydia, dengue, gonorrhea, hepatitis A, influenza, measles, mumps, polio, and whooping cough and plot the predictability ($\chi = 1 - H^p$) as a function of the length of each time series. 
Focusing first on the predictability over short timescales (Figure \ref{fig:smallrambo}), for each time series we average $H^p$ over temporal windows of width up to 100 weeks by selecting 1000 random starting points from each state-level time series for disease and calculating $H^p$ for windows of length 10, 12, ..., 100.
We find that all diseases show a clear decrease in predictability with increasing time window width, which implies that accumulating longer stretches of time series data for a given disease does not translate into improved predictability.  However, we also find strong evidence that the majority of single outbreaks --i.e. temporal horizons characteristic for each disease -- are predictable.  The confidence intervals in Figure \ref{fig:smallrambo} show that there can be large variation in predictability across outbreaks of the same disease, providing a first indication of the presence of a changing underlying model structures and\slash or dynamics on the scale of months. We obtained similar results, e.g., decreasing predictability with time series length, clustering of diseases, and the emergence of barriers to forecasting,  using a weighted version of the permutation entropy, which reduces the dependence of the standard unweighted PE on rare, large fluctuations and by considering estimates of the permutation entropy the time delay, $\tau$, is allowed to vary~\cite{fadlallah2013weighted,garland2014model} (see Supplement). By comparison, across all models with fixed structures studied to date, e.g., white noise, sine waves, and even chaotic systems, the predictability is constant in time or is expected to improve with increasing amounts of time series data~\cite{farmer1987predicting}.

\begin{figure}[H]
\centering
\includegraphics[width=0.99\linewidth]{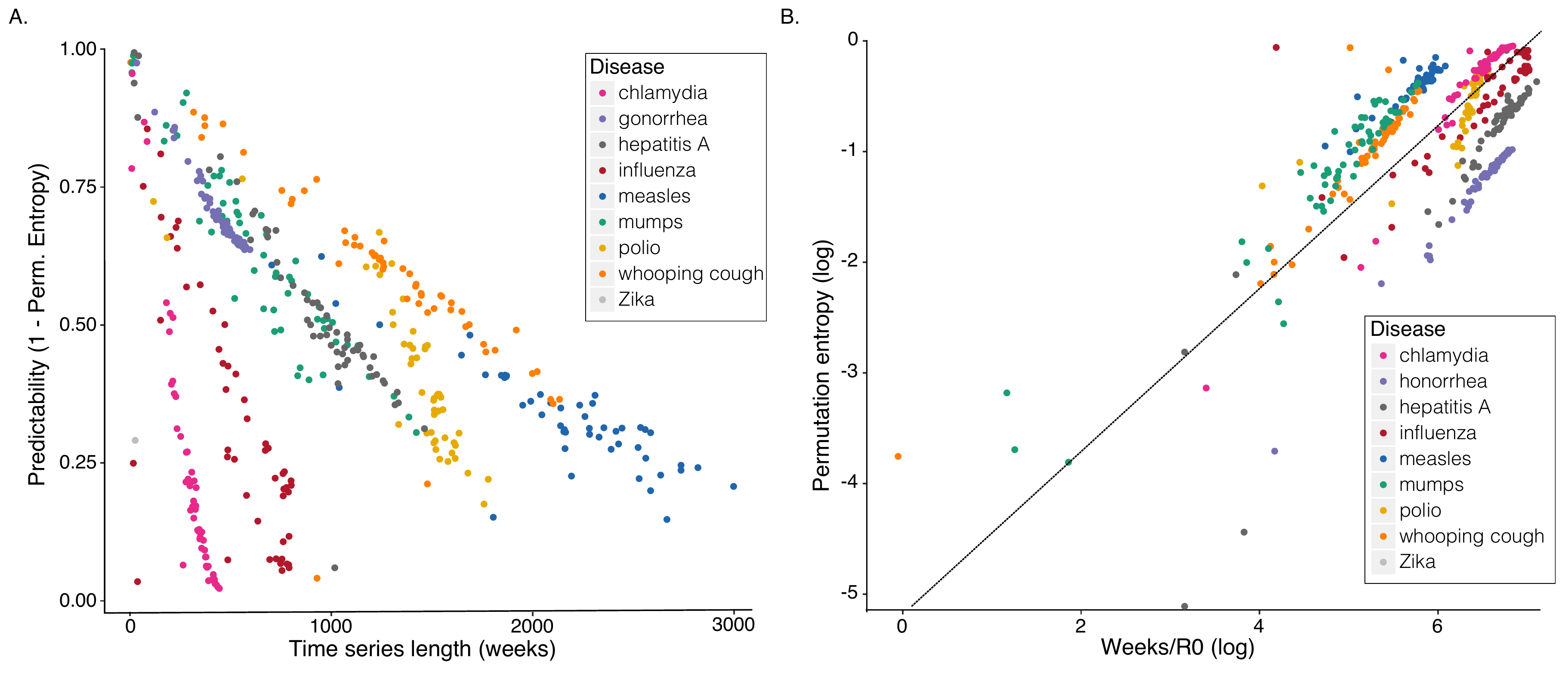}
\caption{\label{fig:bigrambo}\textbf{Permutation entropy and time series length show regularity by disease}.  \textit{A.)} The predictability ($1 - H^p$) for chlamydia, dengue, gonorrhea, hepatitis A, influenza, measles, mumps, polio, and whooping cough is plotted as a function of time series length in weeks.  Although the slopes are different for each disease, in all cases, longer time series, i.e. more data, result in lower predictability.  However, we again find that single outbreaks should be predictable and that diseases show a remarkable degree of clustering based on the slope of entropy gain. In this analysis, each dot represents the predictability for an entire state-level time series for a disease, i.e. the window size is the entire time series. \textit{B.)} Rescaled  time series length based on the mean published basic reproductive number.  Here, we plot the log of time normalized by the basic reproductive number, i.e. $R_0$, from the literature (see Supplement) against the log of the permutation entropy.}
\end{figure}

Zooming out, what is also conspicuous about the relationship between time series length and predictability is that diseases cluster together and show disease-specific slopes, i.e. predictability vs. time series length, which suggests that permutation entropy is indeed detecting temporal features specific to each disease, Figure \ref{fig:bigrambo}A. After re-normalizing time for each disease by its corresponding $R_0$ (the average number of secondary infections a pathogen will generate during an outbreak\slash epidemic when the entire population is susceptible, very large, and is seeded with a single infectious individual)--we used the mean of all reported values found in a literature review (see Supplement)--we find that the best-fit mixed-effect slope on a log scale is one and that the residual effect is well predicted by the times series' embedding dimension $d$ (see Supplement).  Moreover, because the embedding dimension $d$ of a time series is the length of the basic blocks used in the calculation of the permutation entropy, 

\noindent it encodes the fundamental temporal unit of predictability in the form of an entropy production rate, thus implying that predictability decreases with time series data at a disease-specific rate determined to first order by $R_0$, which is further modulated by $d$.  
The result that predictability depends on temporal scale also suggests that the permutation entropy could be an approach for justifying the utility of different data sets, i.e. one could determine the optimal granularity of data by selecting the dimension that maximized predictability. 

One might assume that this phenomenon, i.e. decreasing predictability with increasing time series length, could be driven purely by random walks on the symbolic alphabet used in the permutation entropy estimation. However, $n$-dimensional Markov chain models built from the time series embeddings ($n=d$ the time series' embedding dimension) consistently produced stable and smaller predictability values in comparison with those obtained from data, corroborating that the predictability behaviour we observe does not stem from random fluctuations but is an actual fundamental feature of spreading processes (see Supplement for details on the Markov chain simulations).  This observation, that Markov chain models of the same embedding order do not reproduce the observed predictability, indicates that either the model structure is changing in time and\slash or the system has a very long memory, which is consistent with our current understanding of the entanglement between mobility and disease \cite{szell2011understanding,colizza2007predictability}.  That the best-fit $n$-dimensional Markov chain models over-predict the amount of entropy in real systems, also supports our earlier results that predictable structure does exist across most long outbreak time series. 
 
\begin{figure}[H]
\centering
\includegraphics[width=0.49\linewidth]{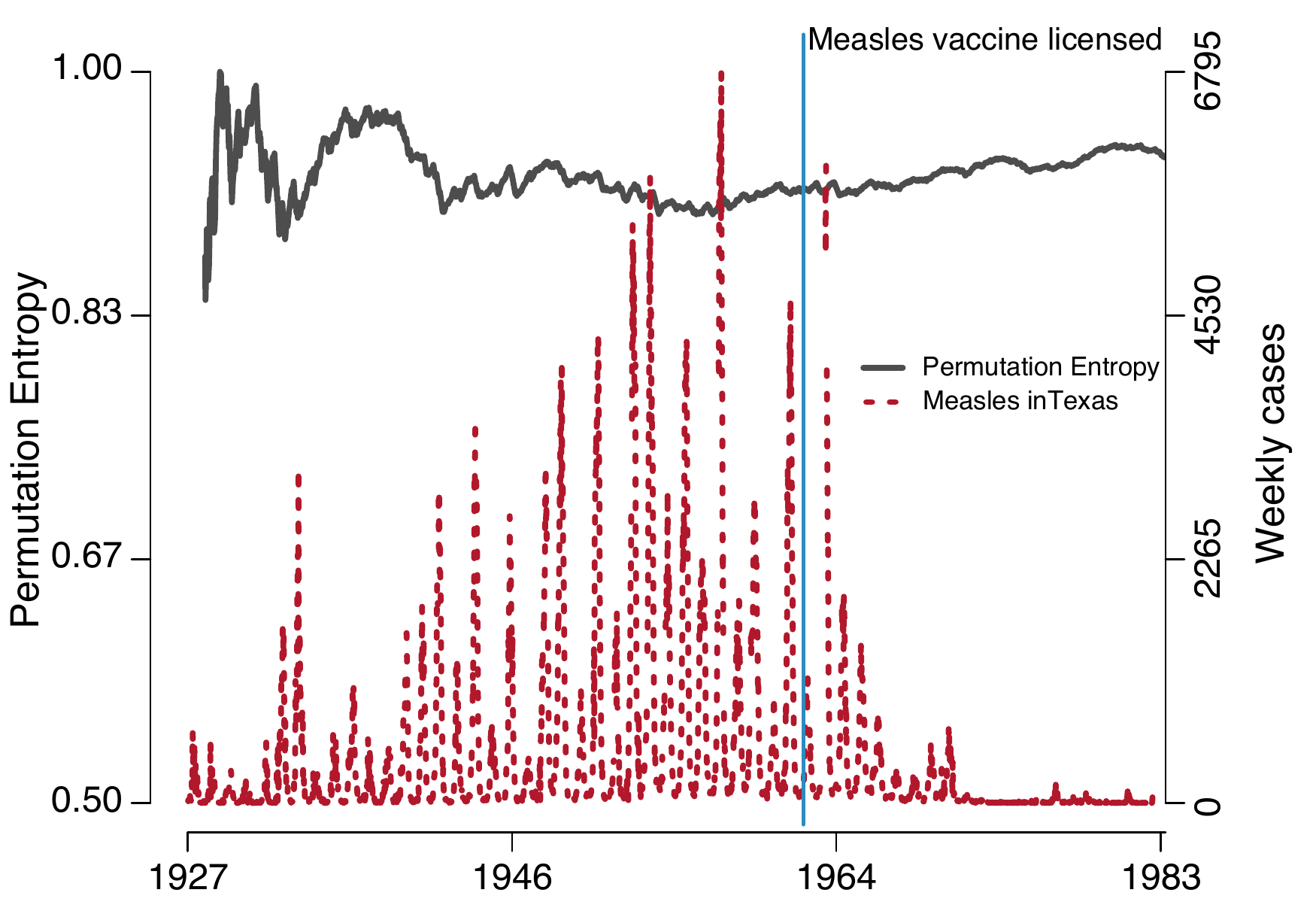}
\includegraphics[width=0.49\linewidth]{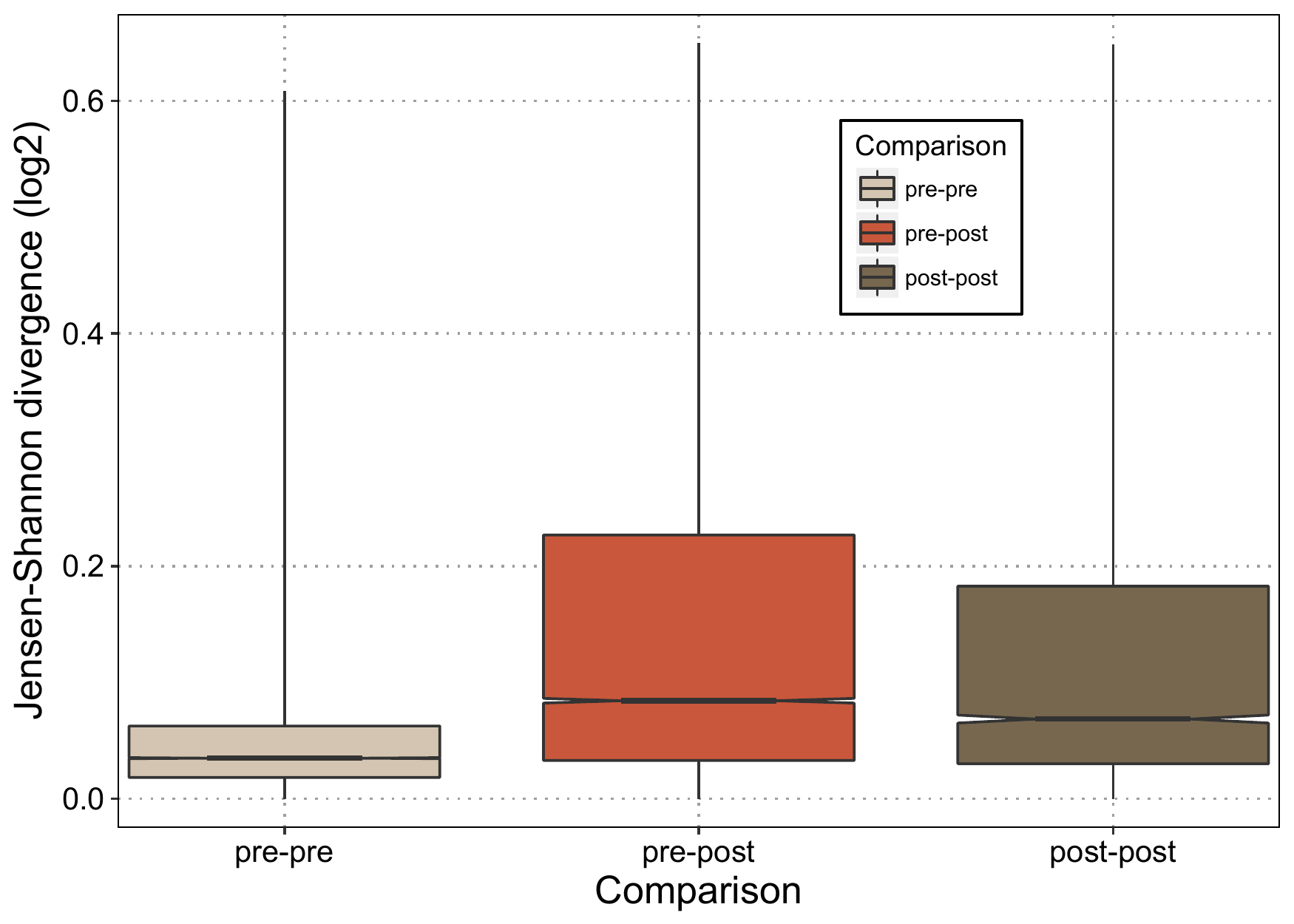}
\caption{\label{fig:vaccent}
\textbf{Permutation entropy shifts after the start of wide-spread measles vaccination.} \textit{A.)} Measles in Texas: permutation entropy shifts after the start of wide-spread vaccination. A rolling window estimate of the permutation entropy for weekly Measles cases reported to the CDC from the state of Texas between 1927 and 1983 (red dashed line).  The vertical blue line indicates when the first measles vaccine was licensed for use in the United States.  Shortly after the vaccine was introduced the permutation entropy increased significantly, which is expected after a system experiences a change to its model structure, in this case vaccination.
\textit{B. )} Permutation entropy detects changing model structure across all states in the US. The box-plots agregate the Jenson-Shannon divergences between pairs of windows that are both in the pre-vaccine era (light brown), have one window in the pre- and one in the post-vaccine era (red) and both in the post-vaccine era (dark drown). The Jenson-Shannon divergence of the symbol frequency distribution was calculated between all pairs of rolling one-year (52 week) windows for measles time series across the US. The average divergence between symbol distributions was higher between windows in the pre- and post-vaccine era, as compared to windows that were either both in the pre- or both in the post-vaccine era (determined by permutation test and by an ANOVA that also included state-level differences in the Jenson Shannon divergence and a Tukey Honest Significant Differences post-hoc test, which contained a correction for multiple comparisons). 
}
\end{figure}

To gain insight into what mechanisms might be driving changes in the predictability, we take advantage of the repeated, ``natural'' experiment of vaccine introduction.  For diseases, such as measles, where we have data from both the pre- and post- vaccine era, we ask whether the permutation entropy changes after the start of wide-spread vaccination.  We consistently observe that predictability decreases after vaccination, again with significance determined by permutation test (see Figure \ref{fig:vaccent}a).  We also find that the symbol frequency distribution changes significantly after vaccination, as measured by the Jenson-Shannon divergence, across all states in the US (see Figure \ref{fig:vaccent}b). Critically, because--as stated earlier--permutation entropy is not affected by changes in magnitude, the difference in entropy cannot simply be accounted for by a reduction in cases.  Instead, it means that the temporal pattern of cases changes after vaccination.  This leads us to the hypothesis that the distribution of secondary infections, its first moment or $R_0$ and its higher moments, drives predictable changes in the permutation entropy, a phenomenon originally discovered in synthetic directed networks by~\citet{meyers2006predicting}.

\begin{figure}[H]
\centering
\includegraphics[width=0.45\linewidth]{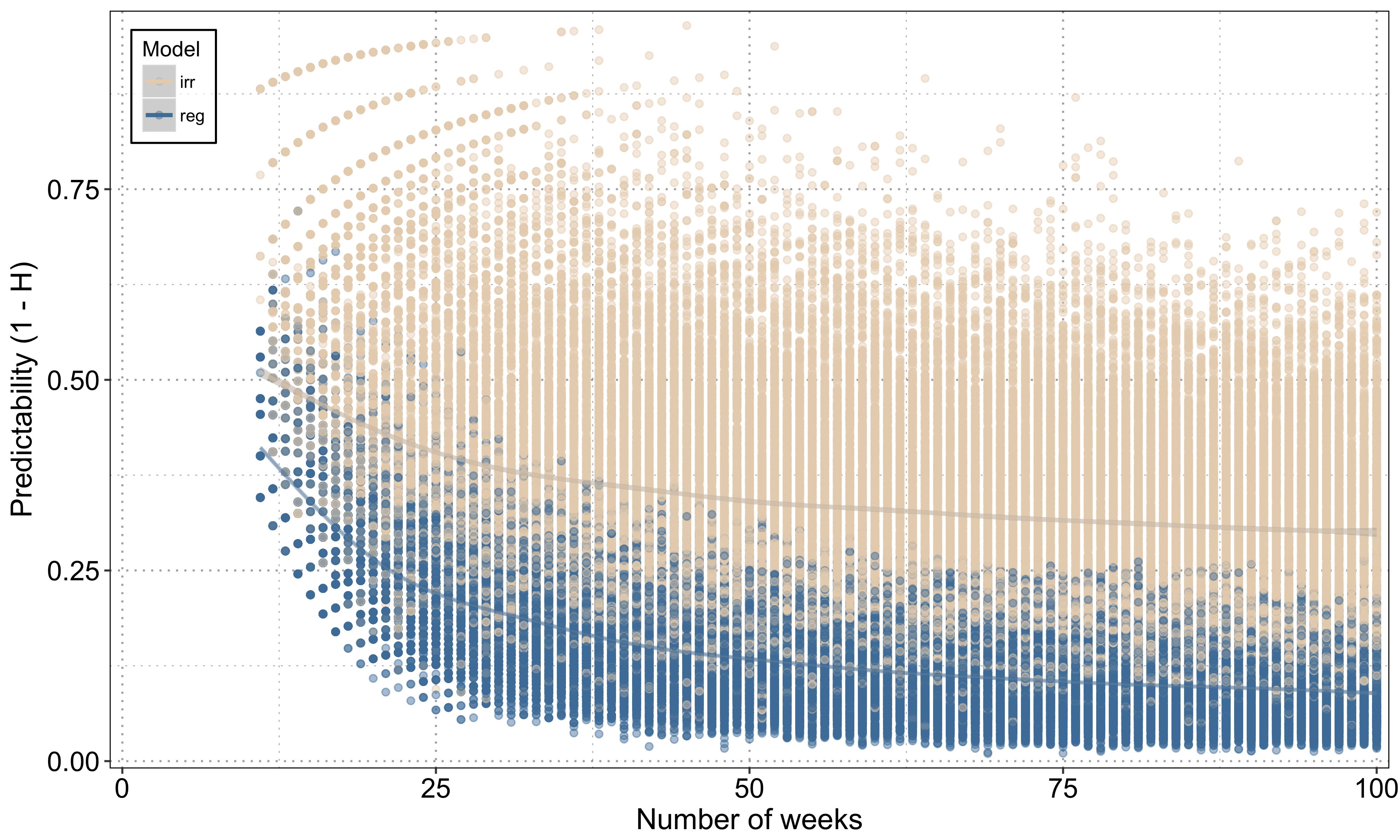}
\includegraphics[width=0.45\linewidth]{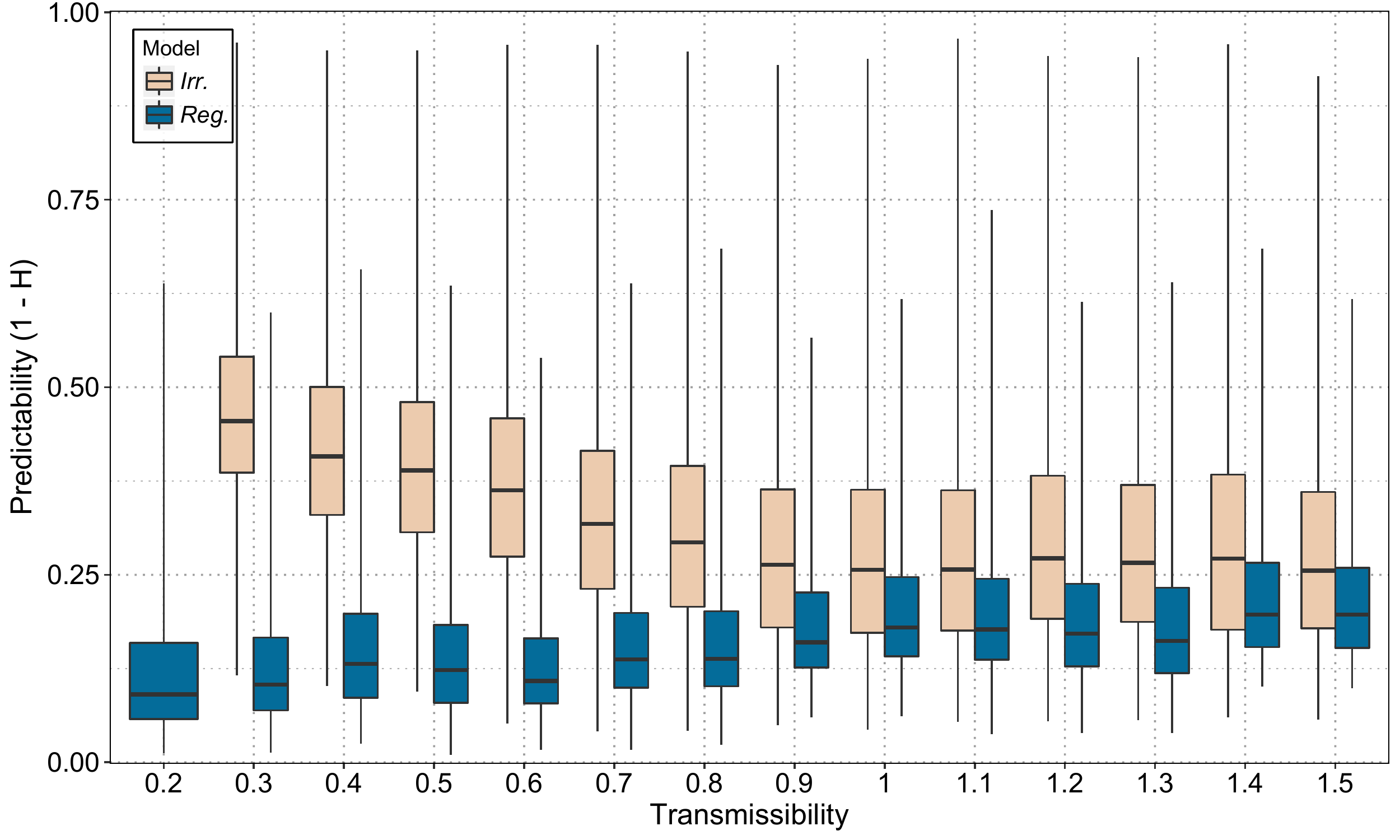}
\caption{\label{fig:simnets}\textbf{Predictability decreases with longer timeseries and contact pattern homogeneity.} 
\textit{A.)} For two classes of contact patterns in a synthetic temporal network--\textit{Reg.}, blue and \textit{Irr.}, brown--we calculate PE for time series of variable length. For each length we chose 1000 random starting points in the time series. We find consistent decreases in the for longer time series in both scenarios. \textit{B.)} When considering the uniform (\textit{Reg.}, blue) and heterogeneous (\textit{Irr.}, brown) scenarios separately, we find that the largest difference in predictability is found below the corresponding epidemic transition consistently with the idea that noisy, sputtering infection trees are harder to predict than the epidemic waves above the transition.
See Supplementary Materials for details.}
\end{figure}

To further evaluate the hypothesis that heterogeneity in the number of secondary infections produces predictable changes in permutation entropy, we simulate an SIR model with probabilistic  restart at end of each outbreak (details in the Supplement) on two classes of temporal networks constructed from the Simplicial Activity Driven (SAD) model \cite{petri2018simplicial}, a modified Activity Driven model in which an activated node contacts $s$ other nodes and induces new links between the contacted nodes (see Supplement for details on the model). 
In this model we can control the epidemic threshold and the number of secondary contacts by changing the activity and the number of contacted nodes per activation. We simulated two scenarios, one in which the number of contacted nodes per activation is fixed (\emph{regular} SAD) and one in which we allow fluctuations in contact number (\emph{irregular} SAD), which generates fluctuations in the number of secondary infections. For both models, we investigated the predictability from below to above the epidemic or critical transmissibility threshold (set to 1 here). 
From the resulting epidemic curves, we calculated the permutation entropy. Figure \ref{fig:simnets}a shows that we find the same pattern of decreasing predictability observed in real data with longer timeseries. 
Figure \ref{fig:simnets}b show the predictability obtained for the two scenarios below and above the transition: we see that that the strongest difference is present below the transition, where the lack of peculiar structure (the regular contact pattern) induces lower predictabilities than for heterogeneous contact distributions. Above the transitions, we find a reduced effect of the difference in contact structure. 

From these results, we can draw three conclusions. 
First, coupled with our earlier results comparing diseases with different average reproductive numbers, heterogeneity in the number of secondary infections can drive differences in predictability, which are related to results on predicting disease arrival time on networks~\cite{shu2012effects} and to recurrent epidemics in hierarchical meta populations~\cite{watts2005multiscale}.  
Second, the permutation entropy could provide a model-free approach for detecting epidemics, which is related to a recent model-based approach based on bifurcation delays~\cite{dibble2016waiting, brett2018anticipating, miller2017forecasting}.  Finally, as outbreaks grow and transition to large-scale epidemics, they should become more predictable, which--as seen in Figures~\ref{fig:timeserie} and~\ref{fig:bigrambo}--appears to be true for real-world diseases as well and agrees with earlier results on how permutation entropy relates to predictability of non-linear systems~\cite{garland2014model}.

Our finding, that horizons exist for infectious disease forecast accuracy and that aggregating over multiple outbreaks can actually decrease predictability is supported by five additional lines of evidence. First, Hufnagel et al. (2004), using data on the 2004 SARS outbreak and airline travel networks, demonstrated that heterogeneity in connectivity can improve predictability~\cite{hufnagel2004forecast}. Second, Domenech de Cell\'es et al. (2018) noted a sharp horizon in forecast accuracy for whooping cough outbreaks in Massachusetts, USA~\cite{de2018impact}. Third, Coletti et al. (2018) demonstrated that seasonal outbreaks of influenza in France often have unique spatiotemporal patterns, some of which cannot be explained by viral strain, climate, nor commuting patterns~\cite{coletti2018shifting}. Fourth, Artois et al. (2018) found that while it was possible to predict the presence human A(H7N9) cases in China, they were unable to derive accurate forecasts for the temporal dynamics of human case counts~\cite{artois2018changing}. Finally, using state-level data from Mexico on measles, mumps, rubella, varicella, scarlet fever and pertussis, Mahmud et al. (2017) showed evidence that while short-term forecasts were often highly accurate, long-term forecast quality quickly degraded ~\cite{mahmud2017comparative}.

Research in dynamical systems over the the past 30 years has demonstrated that prediction error increases with increasing forecast length~\cite{farmer1987predicting}.  However, across that same body work, researchers typically find that predictions improve when they are trained on longer time series, even for chaotic systems~\cite{farmer1987predicting}. Indeed, even for permutation entropy, an active area of research is how spurious aspects of time series can lead to spuriously increasing predictability with increasing time series lengths~\cite{zunino2017permutation}. Our data-driven results suggest that for infectious diseases the opposite is true, more time series data might often lead to lower predictability.  Then, by integrating our biological understanding of each pathogen and simulated outbreaks, we found that changing dynamics, e.g., empirical changes in vaccination coverage and simulated shifts in the number of secondary infections as a disease moves through a heterogeneous social network, can cause the prediction error to increase with increasing data, which is related to earlier findings on the role of airline travel networks and disease forecasting~\cite{colizza2006role}.  What this implies is that different ``models" generate data at different time points and suggests that the optimal coarse-graining of complex systems might change with scale and\slash or time~\cite{wolpert2014optimal}. The potential for scale-dependent models of infectious disease transmission is supported by a recent analysis of US city-level data on influenza outbreaks that found consistent, mechanistic differences in outbreak dynamics based on city size~\cite{Dalziel75}.  

The global community of scientists, public health officials, and medical professionals studying infectious diseases has placed a high value on predicting when and where outbreaks will occur, along with how severe they will be~\cite{hufnagel2004forecast, colizza2007predictability, altizer2013climate, myers2000forecasting}.  Our results demonstrate that outbreaks should be predictable.  However, as outbreaks spread--and spatiotemporally separated waves become entangled with the substrate, human mobility, behavioural changes, pathogen evolution, etc.--the system is driven through a space of diverse model structures, driving down predictability despite increasing time series lengths. Taken together, our results agree with observations that accurate long-range forecasts for complex adaptive systems, e.g., contagions beyond a single outbreak, may be impossible to achieve due to the emergence of entropy barriers.  However, they also support the utility and accuracy of dynamical modeling approaches for infectious disease forecasting, especially those that leverage myriad data streams and are iteratively calibrated as outbreaks evolves.   Lastly, our results also suggest that cross-validation over long infectious disease time series can not guarantee that the correct model for any individual window of time will be favored, which would imply a \textit{no free lunch theorem}~\cite{wolpert1997no} for infectious disease model selection, and perhaps for sociobiological systems more generally.

\subsection*{Data Availability Statement}
Empirical data for all diseases--aside from dengue--were obtained from the U.S.A. National Notifiable Diseases Surveillance System as digitized by
Project Tycho~\cite{tycho}. Dengue data were obtained from the Pandemic Prediction and Forecasting Science and Technology Interagency Working Group under the National Science and Technology Council~\cite{denguedata}.  All code and data associated with this study can be found here: \url{https://github.com/Emergent-Epidemics/infectious_disease_predictability}. The supplement is available at \url{http://scarpino.github.io/files/supplementary-information-predictability.pdf}.

\section*{Acknowledgements}
We thank Joshua Garland and Pej Rohani for productive conversations on permutation entropy and helpful comments on an earlier version of the manuscript. S.V.S. received funding support from the University of Vermont and Northeastern University. G.P. received funding support from Fondazione Compagnia San Paolo. S.V.S. and G.P. conducted performed the study as fellows at IMeRA and drafted the manuscript at Four Corners of the Earth in Burlington Vermont.

\section*{Author contributions}
Both authors conceived the project, performed the simulations and calculations, analyzed the empirical data, interpreted the results, and produced the final manuscript.

\section*{Conflicts of interest}
The authors declare no conflicts of interest exist.


\bibliography{prediction}

\end{document}